# An optical metamixer

Sheng Liu[1†], Polina P. Vabishchevich[1†], Aleksandr Vaskin[2], John L. Reno[1,3], Gordon A. Keeler[1], Michael B. Sinclair[1], Isabelle Staude[2], Igal Brener[1,3,*]

[1]Sandia National Laboratories, Albuquerque, NM 87185, USA

[2]Institute of Applied Physics, Abbe Center of Photonics, Friedrich Schiller University Jena, Albert-Einstein-Str. 15, 07745 Jena, Germany

[3]Center for Integrated Nanotechnologies, Sandia National Laboratories, Albuquerque, NM 87185, USA

*Corresponding author. Email: ibrener@sandia.gov.

†These authors contribute equally to this work.

A frequency mixer is a nonlinear device that combines electromagnetic waves to create waves at new frequencies. Mixers are ubiquitous components in modern radio-frequency technology and are widely used in microwave signal processing. The development of versatile frequency mixers for optical frequencies remains challenging: such devices generally rely on weak nonlinear optical processes and, thus, must satisfy phase matching conditions. In this work, we utilize a GaAs-based dielectric metasurface to demonstrate an optical frequency mixer that concurrently generates eleven new frequencies spanning the ultraviolet to near-infrared (NIR) spectral range. Our approach combines strong intrinsic material nonlinearities, enhanced electromagnetic fields, and relaxed phase-matching requirements, to allow seven different nonlinear optical processes to occur simultaneously. Specifically, when pumped by two femtosecond NIR pulses, we observe second-, third- and fourth-harmonic generation, sum-frequency generation, two-photon absorption induced photoluminescence, four-wave mixing, and six-wave mixing. Such ultracompact optical mixers may enable a plethora of applications in biology, chemistry, sensing, communications and quantum optics.

Mixers are devices that convert electromagnetic wave frequencies and are indispensable in signal processing. For example, radio-frequency mixers have been widely employed in modern communications and navigation systems as modulators, phase detectors, frequency synthesizers, heterodyne receivers, etc. (*1*). Frequency mixers are also in great demand at optical frequencies, where nonlinear crystals are used to generate new colors through nonlinear optical processes, such as harmonic generation, sum- and difference-frequency generation, high-order harmonic generation, and so on. These nonlinear optical processes have greatly broadened the accessible spectrum and are ubiquitous in applications ranging from cutting-edge science and technology to our daily life (such as green color laser pointer). Recent applications of nonlinear optical mixing include attosecond pulse generation (*2*), supercontinuum generation (*3*), optical frequency comb generation (*4*), material characterization (*5*), and quantum optics (*6*). Until now, these applications have relied on bulk nonlinear crystals whose dimensions are much larger than the operating wavelengths and, to achieve efficient frequency conversion, the fundamental and newly generated frequencies need to travel in-phase (i.e. be phase matched) inside the nonlinear medium (6). Consequently, dispersive isotropic materials such as GaAs, although possessing large nonlinear coefficients, cannot be used. Instead, birefringent materials with much smaller nonlinear susceptibilities such as Lithium Niobate and Barium Borate are widely employed. However, due to dispersion, phase matching can only be achieved for one nonlinear process within a narrow bandwidth, and wavelength tuning is achieved by varying incident angles, or using different crystals. Quasi-phase matching (*7-10*) can be used to exploit the large nonlinearities of isotropic materials, but its utility is limited by narrow bandwidth operation, with spectral tuning being achieved by angular rotation, temperature tuning, or electric field bias. Therefore, a device that enables multiple frequency mixing processes across a wide spectral range can be a powerful and versatile platform — however such a device has never been realized using conventional nonlinear optics.

The emergence of resonant metamaterials and metasurfaces has revolutionized our perception of nonlinear optical processes. In contrast to bulk nonlinear optical crystals, subwavelength resonant cavities (*11-14*) greatly enhance electromagnetic fields in tight volumes (*15*) and relax phase matching conditions. This allows the simultaneous occurrence of various nonlinear processes (*16-18*). More recently, semiconductor dielectric metamaterials, operating below the bandgap, have attracted intense attention due to their low material losses at optical frequencies (*19, 20*) as well as their resonant interaction with both the electric and magnetic fields (*21*). In particular, free-carrier effects (*22*), two-photon-absorption (*23*), second- (*24*) and third-harmonic generation (SHG and THG) (*25*) as well as sum-frequency generation (SFG) (18) were studied in silicon or III-V semiconductor nanoresonators or their periodic arrangements. However, frequency conversion beyond second- and third-order nonlinear processes has never been observed using either metallic metamaterials or dielectric metamaterials.

Here, we demonstrate an optical metamixer – a GaAs-based dielectric metasurface that enables a variety of simultaneous nonlinear optical processes across a broad spectral range. Specifically, seven different nonlinear processes (second-, third- and fourth-harmonic generation, sum-frequency generation, two-photon absorption induced photoluminescence, four-wave mixing and six-wave mixing) simultaneously give rise to eleven new frequencies that span the ultraviolet to NIR spectral range. Our multifunctional metamixer exploits the combined attributes of resonantly enhanced electromagnetic fields at the metasurface resonant frequencies; large even- and odd-order optical nonlinearities of GaAs; and significantly relaxed phase matching conditions due to the subwavelength dimensions of the metasurface.

Fig. 1A shows a schematic of nonlinear frequency generation by a GaAs metasurface pumped by two laser beams. The left inset of Fig. 1A shows a 60° side-view scanning electron microscope (SEM) image of a typical GaAs metasurface used in these measurements. The metasurface consists of a periodic square array of nanocylinders with a diameter of ~400 nm. Each nanocylinder consists of three layers: the top $SiO_x$ etch mask (~200 nm), the middle ~400 nm thick GaAs nanodisk that confines the electromagnetic field, and the bottom low refractive index $(Al_xGa_{1-x})_2O_3$ layer (~450 nm) for isolating the GaAs nanocylinder from the high index GaAs substrate. The resonantly enhanced frequency mixing is achieved by exciting the lowest order magnetic dipole (MD) and electric dipole (ED) resonances of the GaAs nanocylinder (*19, 26, 27*) simultaneously. The measured reflectivity spectrum (right inset of Fig. 1A) exhibits maxima at $\lambda_1 \sim 1270$ nm and $\lambda_2 \sim 1520$ nm which correspond to the ED and MD resonances, as identified by the simulated electric field profiles shown in the insets.

First, we study harmonic generation by the GaAs metasurface when pumped by a single near-IR femtosecond beam with a wavelength near the MD resonance ($\lambda_1$~1570 nm) using an average power of ~4.5 μW. We used a 20X near-IR objective with a numerical aperture NA= 0.4 to both focus the pump pulses on the GaAs metasurface and collect the harmonic beams. Fig. 1B shows the second-, third- and, fourth-harmonics (inset of Fig. 1B) generated by the metasurface at $\lambda_{SHG} \sim 785$ nm, $\lambda_{THG} \sim 523$ nm, $\lambda_{FHG} \sim 393$ nm, respectively. The efficiency of the SHG process is estimated to be $2.3 \times 10^{-6}$. The emission centered at $\lambda_{PL} \sim 870$ nm corresponds to GaAs photoluminescence (PL) arising from two-photon absorption of the pump (*28*). Note that the observed harmonics are above the GaAs bandgap energy and therefore suffer from significant material absorption.

Next, we introduce a second femtosecond pump beam spectrally tuned to $\lambda_2 \sim 1240$ nm to overlap the resonators' ED mode. The collinearly propagating pump beams were focused at the same location on the sample with average powers of $P_1 \sim 3.6$ μW and $P_2 \sim 5$ μW, respectively. When the two pump beams are temporally coincident, we observe eleven spectral peaks, ranging from ~380 nm to ~1000 nm (Figure 2A).

We categorize the generated signals into two groups. The first group, indicated by the blue labels, corresponds to harmonic generation processes and also PL arising from two-photon absorption. Each of the processes in this group relies on only a single pump beam. In contrast, the second group, indicated by red labels, corresponds to frequency mixing processes that require both pump pulses. The five frequency mixing signals include: sum-frequency generation ($\omega_1 + \omega_2$) at $\lambda_{SFG} \sim 689$ nm; three types of four-wave mixing (FWM) ($2\omega_2 - \omega_1$, $2\omega_1 + \omega_2$, $2\omega_2 + \omega_1$) at ~1000 nm, ~472 nm, and ~434 nm, respectively; and six-wave mixing (SWM) ($4\omega_1 - \omega_2$) at $\lambda_{SWM} \sim 577$ nm. Note, that the amplitude of the FWM ($2\omega_2 - \omega_1$) peak is at least ten times higher than any of other nonlinear processes. This is attributed to the much lower absorption below the GaAs bandgap in contrast to the strong attenuation the other signals experience. Altogether, the measured spectra contains simultaneous contributions arising from seven nonlinear optical processes.

We verify the physical origin of the frequency mixing processes by measuring the output spectra for various pump wavelengths and for different pump powers. For example, the SFG output is identified due to its energy coinciding with the sum of the photon energies of the two pump beams $\omega_{SFG} = \omega_1 + \omega_2$, as well as by the linear intensity dependence on one pump power when the other pump power is held constant (Fig. 2B black curve). In a similar manner, we performed power dependence measurements to verify the four-wave and six-wave mixing processes. The red curve in Fig. 2B shows the quadratic dependence of the FWM ($2\omega_2 - \omega_1$) output on the $\omega_2$ pump power, and Fig. 2C shows a linear dependence of the SWM intensity on the power of the pump at $\omega_2$. To further confirm the SWM process, we spectrally tuned both pump wavelengths and observed excellent agreement between the measured and calculated SWM peak locations (Fig. 2D). To confirm that the observed nonlinear processes are enhanced by pumping at the dipolar Mie resonances, we performed similar frequency mixing measurements on both the unpatterned GaAs substrate and other GaAs metasurfaces with different diameter resonators, and, as expected, significantly lower signal intensities were observed.

To investigate the temporal dynamics of the nonlinear generation processes, we measured the signal intensities while varying the optical delay between the two pump pulses. Fig. 3 shows a 2D contour image of the transient nonlinear conversion. As expected, the harmonic generation signals and PL arising from two-photon absorption which each requires only one of the pumps are observed regardless of the optical delay. In contrast, the frequency mixing signals such as SFG, FWM, and SWM appear only within a 160 fs window when the two pump pulses temporally overlap at the metasurface. Note that we also observe time-dependent spectral shifts of the SFG and FWM signals. This is likely due to the chirp of the pump pulse: for example the negative chirp of $\omega_2$ that the higher frequencies arrive at the metasurface earlier than the lower frequencies.

The nonlinear generated signals shown in Figures 1-3 were measured when the far-field collection angle was optimized for the SFG intensity, and we observe significant increases of the other signal intensities when they are optimized individually. This indicates a variety of far-field emission profiles for different nonlinear signals, which significantly limits the measured conversion efficiencies, especially considering the low NA objective used (*29*). Moreover, the majority of the nonlinearly generated frequencies lie above the GaAs bandgap and thus experience intense absorption. The conversion efficiencies can be improved by using larger bandgap materials such as AlGaAs to reduce absorption loss, or by fabricating resonators with larger dimensions so the Mie resonances shift to longer wavelengths. Dielectric resonators are also a powerful platform that provides microscopic control over the electric field intensity and polarization distributions inside the resonator. Therefore, higher conversion efficiencies can be expected by engineering the resonator shape to optimize the modal overlap between the pump frequencies and nonlinear generated frequencies. Moreover, III-V metasurfaces could enable tailoring of the polarization and far-field profiles of the nonlinear emission (*11, 14*) beyond what is possible with a single resonant particle (*29*). Finally, the use of lower pump power might be possible when using dielectric metasurfaces based on high quality-factor resonators which exhibit much larger electromagnetic field enhancements (*30, 31*).

Our experimental demonstration of seven different nonlinear optical processes occurring simultaneously generated in GaAs metasurfaces could provide the opportunity for realizing ultracompact optical mixers for various applications. The observed high order nonlinear processes might allow generation of high-order harmonics which is the foundation of attosecond pulse generation (*32*). Compact nonlinear optical mixers could be used in optical telecommunication devices. Moreover, we anticipate that these metasurfaces could be optimized for other nonlinear mixing processes such as difference frequency generation $\omega_{DFG} = \omega_1 - \omega_2$. This would enable the production of femtosecond pulses covering the Mid-IR spectral range, where broadband laser-gain media and saturable absorbers do not exist.

**Acknowledgments:** Parts of this work were supported by the U.S. Department of Energy, Office of Basic Energy Sciences, Division of Materials Sciences and Engineering and performed, in part, at the Center for Integrated Nanotechnologies, an Office of Science User Facility operated for the U.S. Department of Energy (DOE) Office of Science. Sandia National Laboratories is a multi mission laboratory managed and operated by National Technology and Engineering Solutions of Sandia, LLC, a wholly owned subsidiary of Honeywell International, Inc., for the U.S. Department of Energy's National Nuclear Security Administration under contract DE-NA0003525. A.V. and I.S. gratefully acknowledge financial support from the German Research Foundation (STA 1426/2-1) and by the Thuringian State Government through its ProExcellence Initiative ($ACP^{2020}$).

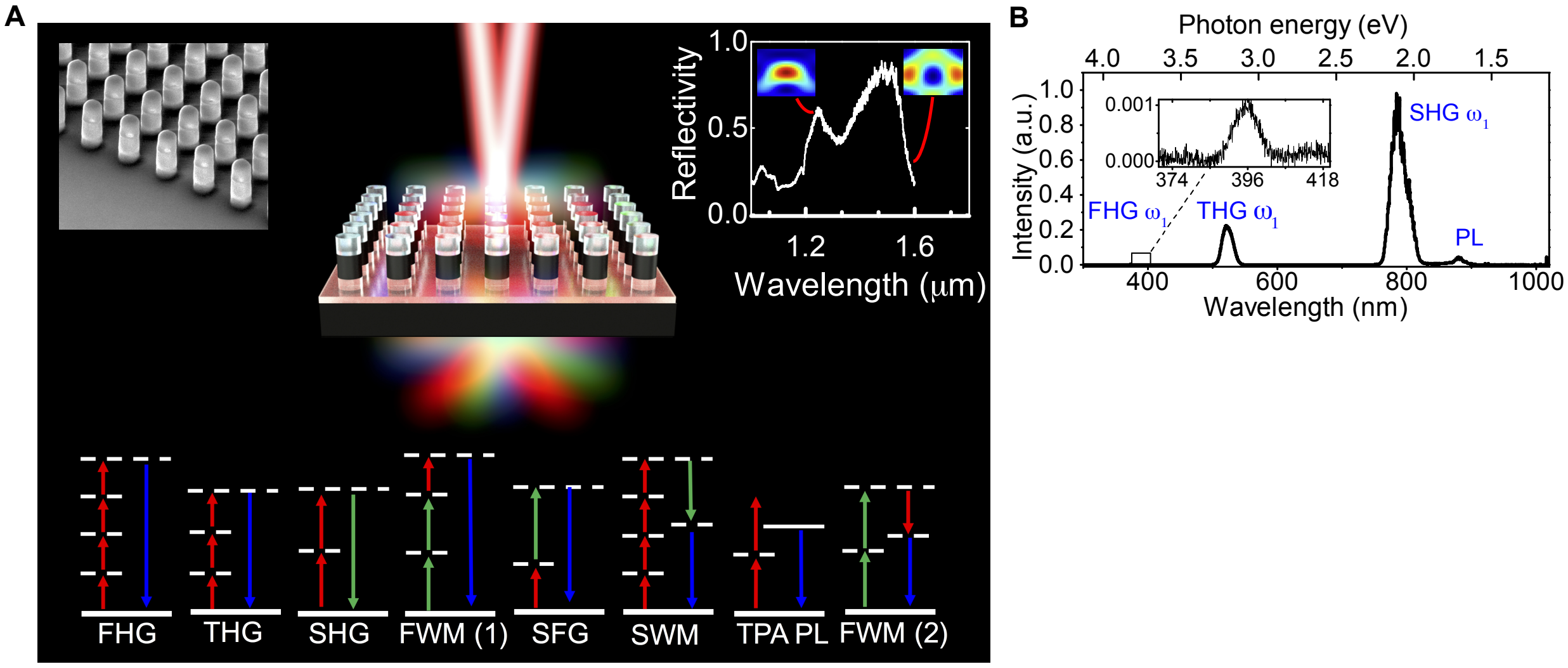

**Fig. 1.** (**A**) The schematic of an optical metamixer consisting of a square array of subwavelength GaAs dielectric resonators. Two femtosecond near-IR pulses pump the metamixer and a variety of new-frequencies are simultaneously generated. Top left inset: a 60°side-view scanning electron microscope image of the GaAs metamixer. Top right inset: the reflectance spectrum of the metasurface with two cross-section local field distributions corresponding to the electric and magnetic dipole overlapping the center wavelengths of the pump beams. Bottom inset: schematic energy diagrams of the seven nonlinear optical processes that occur simultaneously in our metamixer. **(B**) Spectra of second-, third- and fourth-harmonics when pump pulses of $\lambda_1$~1570 nm are used to excite the GaAs metamixer. Inset is the zoom-in of the fourth harmonic generation.

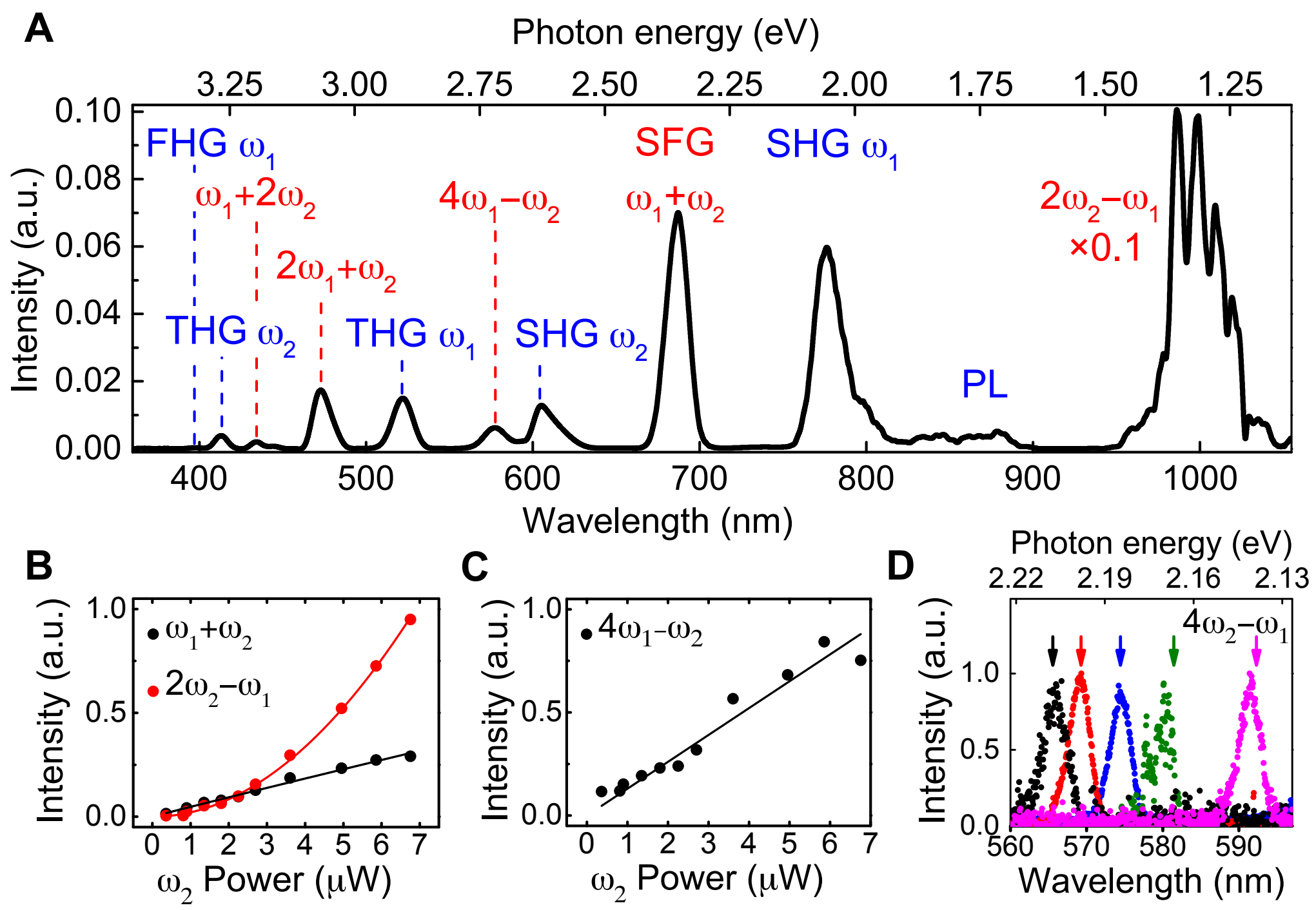

**Fig. 2.** (**A**) Spectrum exhibiting eleven nonlinearly generated peaks originating from seven different nonlinear processes when two optical beams at $\lambda_1 \sim 1240$ nm and $\lambda_2 \sim 1570$ nm are used to simultaneously pump the GaAs metasurface. Blue labels indicate harmonic generation processes and photoluminescence arising from two-photon absorption that each requires only one pump beam. Red labels indicate frequency mixings that involves both pump beams. (**B** and **C**) Dependence of the sum-frequency generation $(\omega_1 + \omega_2)$, four-wave mixing $(2\omega_2 - \omega_1)$, and six-wave mixing $(4\omega_1 - \omega_2)$ intensities on the power of the $\omega_2$ pump. Both the experimental data (dots) and theoretical fitting (black line for linear fitting and red curve for quadratic fitting) are shown. (**D**) Five representative spectra showing the tuning of the normalized six-wave mixing $(4\omega_1 - \omega_2)$ signal when the pump wavelengths are spectrally tuned to $\lambda_1$~1248.7 nm, $\lambda_2$~1557.5 nm (black curve); $\lambda_1$~1234.8 nm, $\lambda_2$~1558.6 nm (red curve); $\lambda_1$~1211.6 nm, $\lambda_2$~1558.9 nm (blue curve); $\lambda_1$~1234.9 nm, $\lambda_2$~1581.2 nm (green curve); and $\lambda_1$~1233.7 nm, $\lambda_2$~1600.4 nm (magenta curve). The arrows denote the theoretically expected frequencies for the considered six-wave mixing process.

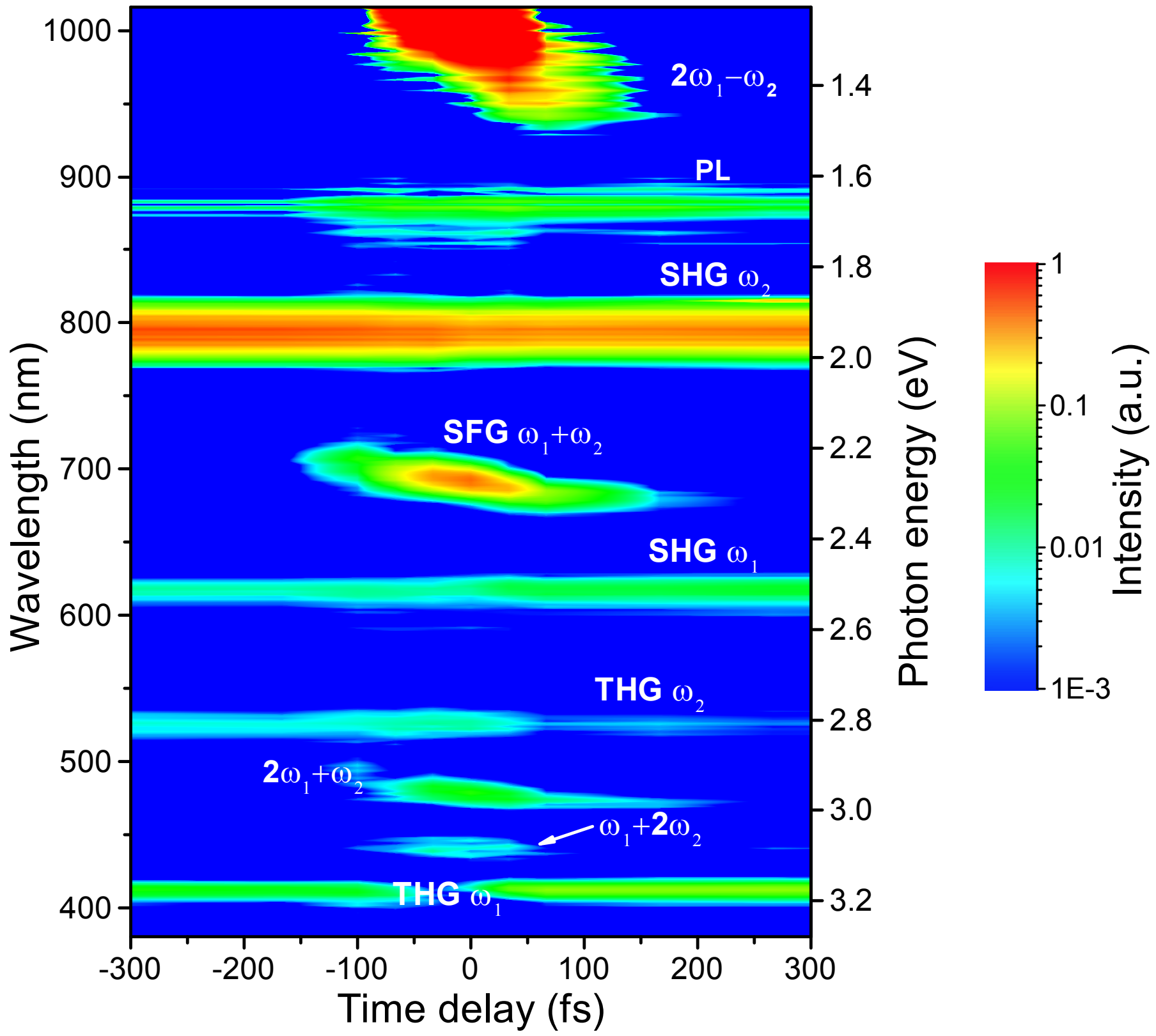

**Fig. 3.** 2D contour image of the transient nonlinear signal (logarithmic scale) when the time delay between the two pump pulses is varied. The nonlinear signals that require only one of the pumps do not depend on the delay, while the mixing signals that rely on both pumps occur only when the two pumps overlap in time.